\documentclass[aps,prb,nobalancelastpage,twocolumn]{revtex4}
\usepackage[dvips]{graphicx}

\begin{document}

\title{Ferrodistorsive orbital ordering in the layered nickelate
NaNiO$_{2}$: A density-functional study}

\author{H. Meskine and S. Satpathy}

\date{\today}

\begin{abstract}
The electronic structure and magnetism in the sodium nickelate NaNiO$_{2}$ in the
low-temperature phase is studied from density-functional calculations using the
linear muffin-tin orbitals method. An
antiferromagnetic solution with a magnetic moment of 0.7 $\mu_{B}$ per Ni
ion is found. A ferrodistorsive orbital ordering is shown to occur due to the 
Jahn-Teller distortion around the Ni$^{+3}$ ion in agreement with the orbital ordering
inferred from 
neutron diffraction. While the intralayer exchange is ferromagnetic, the interlayer
exchange is weakly antiferromagnetic, mediated by a long Ni-O-Na-O-Ni superexchange path.
\end{abstract} 
\maketitle

Since their discovery in the early fifties\cite{Dyer-JAmChemSoc:1},
the nickelates have been the subject of attention due to their many
applications such as the base materials in batteries. More recently, the
family of compounds Li$_{x}$Me$_{1-x}$NiO$_{2}$ (Me being a metal)
have attracted special interest due to their triangular lattice
structure which makes them good candidates as frustrated magnetic systems.
In fact, magnetic or orbital frustration mechanisms are usually invoked
in explaining the unusual behavior of LiNiO$_{2}$. Many different
scenarios have been proposed to explain the absence of orbital and
magnetic orderings in LiNiO$_{2}$, such as spin-glass, \cite{Bajpai:prb1}
quantum disordered state,\cite{li:3527,kitaoka:jpsj} spin-orbital
liquid,\cite{Feiner:prl1} impurity effect,\cite{mostovoy:227203}
or frustrated anti-ferromagnet.\cite{yoshizawa:jpsj} In comparison, NaNiO$_{2}$
seems to show none of the unusual behaviors of its sister 
compound including the magnetic behavior.\cite{Chappel:EPB:2,Chappel:EPB:3}
For example, while in LiNiO$_{2}$,  no
long-range magnetic order has been observed, NaNiO$_{2}$ in contrast
is a type A antiferromagnet
(ferromagnetic layers stacked on top of one another and coupled antiferromagnetically). 
It has been a long puzzle as to why their
magnetic properties are so different, in spite of the fact that the
 two compounds are very similar. 

In this paper, we study the electronic structure of
NaNiO$_{2}$ using  density-functional methods, with the goal of gaining insight into the
physics of this family of materials. The nature of orbital ordering as well as the origin
of the magnetic exchange are also discussed.

\begin{figure}[!tp]
\begin{center}\includegraphics[%
  width=5.5cm]{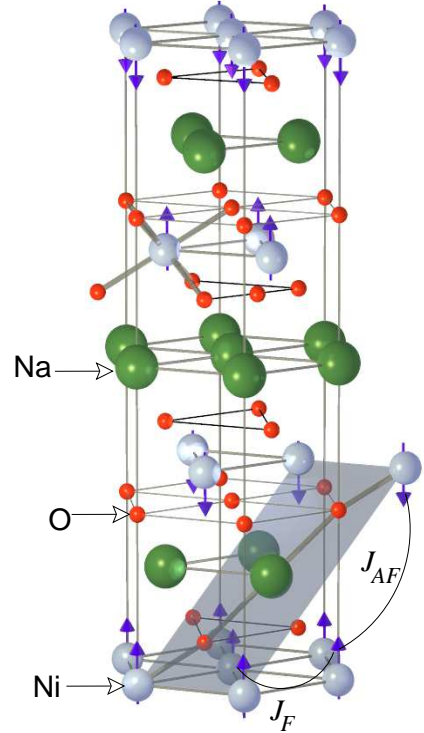}
\end{center}

\caption{ High temperature crystal structure of 
NaNiO$_{2}$. Calculations reported here are for the
low-temperature phase ($C2/m$), but the high-temperature structure ($R\bar{3}m$) is shown
here because  it is simpler to visualize. The former is obtained by
distorting the latter, including a
Jahn-Teller stretching of the 
NiO$_{6}$ octahedra.   The magnetic
ordering is antiferromagnetic type A, with ferromagnetic interaction within the
layer and antiferromagnetic interaction between the layers. The line
connecting the Ni-O-Na-O-Ni  atoms indicates the superexchange path and the
rectangle joining the four Ni atoms has reference to the plane of the contour plot in Fig.
4.
\label{fig:crystal structure}}
\end{figure}

The high-temperature hexagonal crystal structure of NaNiO$_{2}$ (space group
$R\bar{3}m$, no. 166) is shown in Fig. \ref{fig:crystal structure}. The compound
undergoes a structural transition to a lower-symmetry monoclinic structure
with the paramagnetic space group $C2/m$ (no. 12) at about 450 K,
below which the oxygen octahedra become elongated. 
The magnetic transition is at a much lower temperature $T_N = 20 K$\cite{Chappel:EPB:3},
below which the A-type antiferromagnetic structure occurs.
The layered structure may be viewed as an
arrangement of slightly elongated NiO$_{6}$ octahedra separated by Na sheets.
In the low-temperature structure, the Jahn-Teller (JT) distortion leads to 
  different Ni-O
bond lengths: four short bonds of  1.91 \AA and two long ones of 2.14 \AA.
The lattice parameters are listed in Table \ref{tab:atom positions}.
In the low-temperature phase, NaNiO$_{2}$ shows the ferro-distorsive
orbital ordering caused by the Jahn-Teller distortions of the
NiO$_6$ octahedra, which disappears in the high-temperature phase.

\begin{figure}[!tp]
\begin{center}\includegraphics[%
  width=6.5cm,
  keepaspectratio]{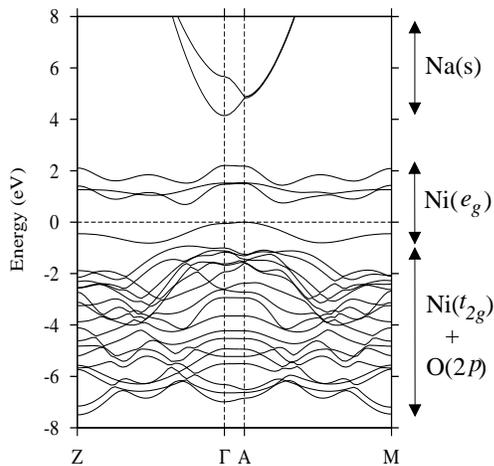}
\end{center}

\caption{Density-functional energy bands for the antiferromagnetic  NaNiO$_{2}$ in the
low-temperature structure obtained from the ``LDA+U" calculations.
The unit cell has two formula units: ( NaNiO$_{2}$)$_2$.
The 
$e_{g}$ bands are split near the Fermi level due to the Jahn-Teller and exchange
interactions.
\label{fig:Energy-Bands}}
\end{figure}
The band-structure calculations were performed for the low temperature
structure using the local spin-density approximation (LSDA) to the density
functional theory (DFT) as well as the ``LDA+U" approximation, using the linear
muffin-tin orbitals (LMTO) method\cite{andersen:2571,andersen:3060}.
For the antiferromagnetic structure, the symmetry is not reduced further and
the magnetic unit cell is also monoclinic with space group $C2/m$
with two formula units per unit cell. Within the LMTO atomic sphere
approximation (LMTO-ASA), the antiferromagnetic (AF) type A structure
was found to be the ground state, lower in energy than both the ferromagnetic
(FM) and paramagnetic structures.

In the ``LDA+U" method,\cite{LDA-Review} the Coulomb interactions between the localized
electrons (Ni(d)) are subtracted from the LDA total energy and a ``more proper"
Hubbard-like term is added, so that the corrected energy functional reads:

\begin{equation}
E=E_{LDA}-UN(N-1/2)+(U/2)\sum_{i\ne j} n_i n_j,
\end{equation}
where N is the total number of the localized electrons, $n_i$ is the occupancy
of the $i^{th}$ localized orbital, and U is the screened Coulomb energy. The orbital
energies are obtained by taking the derivatives with respect to the orbital occupation, so
that

\begin{equation}
\epsilon_i=\partial E/ \partial n_i = \epsilon_{LDA}+ U (1/2-n_i).
\end{equation}
Thus with respect to the LDA results, the energies of the occupied orbitals are shifted 
by $-U/2$, while those of the unoccupied states are shifted up by the amount $U/2$.
In the actual calculations, one works with a slightly more complicated functional that
includes properly the direct and the exchange Coulomb interactions, parametrized by
the screened interactions $U$ and $J$. 

The band structure of NaNiO$_{2}$ in the AF-type A magnetic structure
calculated with the ``LDA+U" method, with $U=5$ eV and $J=0.5$ eV, are shown in Fig.
\ref{fig:Energy-Bands}. While the LSDA  results showed a metallic band structure because
of a slight overlap between the bands derived from the $e_g^1\uparrow$ and
$e_g^1\downarrow$ Ni(d) states at the same Ni site, the overlap disappears when the
correlation effects are included within the   ``LDA+U" formalism. Only a small value of U
is needed to open up the gap, suggesting a robust insulating gap in the compound resulting
from electronic correlation. 

\begin{figure}[h]
\begin{center}\includegraphics[%
  width=6.5cm,
  keepaspectratio]{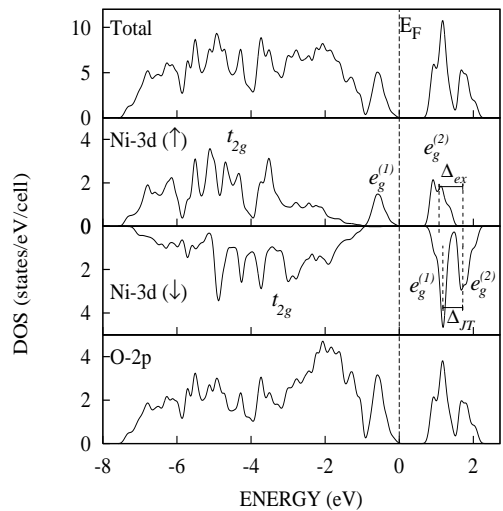}
\end{center}

\caption{Electronic Density of states for NaNiO$_{2}$. 
$\Delta_{JT}$ and $\Delta_{ex}$ refer to the Jahn-Teller
and the exchange splittings, respectively. 
\label{fig:density of states}}
\end{figure}

The bands indicate a low-spin state , with the $t_{2g}$
states being completely occupied while the $e_{g}$ states are only
quarter-filled ($t_{2g}^{6}e_{g}^{1}$). As indicated from the densities-of-states
(Fig. \ref{fig:density of states}), the $t_{2g}$ and $e_{g}$
bands are split by a strong crystal-field splitting, while the $e_{g}^{\uparrow}$ and
$e_{g}^{\downarrow}$ are split by the exchange coupling by the amount $\Delta_{ex}\approx$
0.7 eV. The Ni($e_g^{1}$) occupancy makes the system JT-active in the low-temperature
phase, lifting the degeneracy of the $e_{g}$ levels,
with a JT energy of $\Delta_{JT}\approx$0.6 eV.

\begin{table}[t]

\caption{Atomic positions and sphere radii used in the LMTO calculations.
Lattice parameters for the monoclinic unit cell (space group $C2/m$)
are $a=5.311$ \AA, $b=2.840$ \AA, $c=5.568$\AA, and
$\beta=110.44^{\circ}$ (Ref. \cite{Chappel:EPB:3}).\label{tab:atom positions}}

\begin{center}\begin{tabular}{cccccc}
\hline 
atom&$x/a$&$y/b$&$z/c$&$S$ {[} \AA{]}&site
\tabularnewline
\hline
Na&0&1/2&1/2&1.72&2d
\tabularnewline
Ni&0&0&0&1.28&2a
\tabularnewline
O&0.2832&0&0.804&1.04&4i
\tabularnewline
\hline
\end{tabular}
\end{center}
\end{table}

\begin{figure}[!tp]
\begin{center}\includegraphics[%
  width=6.5cm, 
  keepaspectratio]{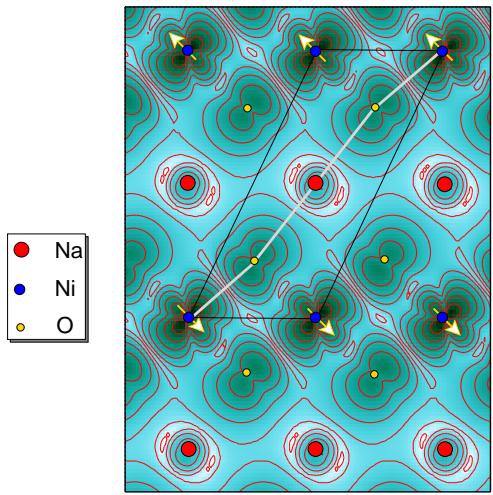}
\end{center}
\caption{ (Color online) Charge density contours of the occupied 
Ni (d) bands (about 0.8 eV wide occurring just below $E_F$) showing the ferro-orbital 
ordering of the Ni($e_{g}$) orbitals, as obtained from the LSDA calculations.
The  plane of the plot is the diagonal plane, defined by the rectangle
drawn here and in Fig.
\ref{fig:crystal structure}. Arrows indicate the  spins of
the Ni$^{3+}$ ions and the line joining the Ni-O-Na-O-Ni indicates
the superexchange path. Contour values
are given by
 $\rho_{n}=\rho_{0}\times10^{-n\delta}$,
with $\rho_{0}=0.708$ $e^{-}/a_{0}^{3}$, $\delta=0.575$, 
and $n=0,...,7.$
\label{fig:charge density}}
\end{figure}
The calculated magnetic moment is $\mu=$0.51 $\mu_{B}$/Mn obtained from the LSDA,
which increases to 0.71 $\mu_{B}$/Mn  in the ``LDA+U" results, as in the
latter, the slight overlap between  the $e_g^1\uparrow$ and $e_g^1\downarrow$ disappears,
increasing the occupancy of the spin $\uparrow$ states.
This is lower than the nominal magnetic moment of 1 $\mu_{B}$/Ni because of the strong
hybridization between the O(2$p$) and the Ni(3$d$) states.

We have studied the orbital ordering by examining the occupancy of the
various Mn (d) orbitals.  In a frame of reference where the $z$ axis points along the 
long Ni-O bond,
we found the occupied Ni(d) electrons to be predominantly of $3z^{2}-r^{2}$ character. 
The charge density contours for the occupied Mn (d) states, plotted in Fig.
\ref{fig:charge density}, shows the ferro-orbital ordering of the 
$3z^{2}-r^{2}$ orbitals, with significant hybridization coming from the O (2p) states. 
Fig. 5 shows a sketch of the orbital order as obtained from the density-functional
charge density. The charge density for the undistorted high temperature structure was also
computed and  we found no orbital order for that case.

The magnetic exchange interactions, the ferromagnetic  $J_F$ within the layer
and the antiferromagnetic $J_{AF}$ between the layers are both weak, with the measured
values being $J_F = 13$ K and $J_{AF}=-1$ K.\cite{Chappel:EPB:2}
The exchange interactions are weak because the {\it intra}layer exchange involves 
a 90$^{\circ}$ Ni-O-Ni bond, while for the {\it inter}layer coupling, it is the  long
Ni-O-Na-O-Ni superexchange path that makes the exchange weak. 
The 90$^{\circ}$ Ni-O-Ni exchange within the layer has been examined in great detail  
in two recent papers\cite{Khomskii,Euro}. 
\begin{figure}[!tp]
\begin{center}\includegraphics[%
  width=5.0cm, 
  keepaspectratio]{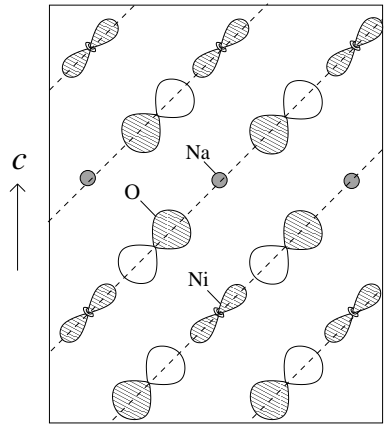}
\end{center}
\caption{ Sketch of the orbital order in NaNiO$_2$ extracted from Fig. 4. The shaded lobes
indicate a positive wave function, indicating the anti-bonding interaction between the Ni
(3$z^2-r^2$) and the O(p$_z$) orbitals.
\label{fig:OrbitalStructure}}
\end{figure}

For the exchange between the layers,
we  treat
the Ni-O-Na hopping as an effective Ni-Na hopping between the Ni($3z^{2}-r^{2}$)
and Na($s$) orbitals on the Ni-O-Na-O-Ni superexchange path. 
The model Hamiltonian for the effective Ni-Na-Ni path then
reads
\begin{equation}
H_{el} = \sum_{ij,\sigma}t\ 
c_{i\sigma}^{\dagger}c_{j\sigma}+
h. c. +\sum_{i\sigma}\varepsilon_{i}n_{i\sigma}
 +  \sum_{i}U_{i}n_{i\uparrow}n_{i\downarrow},
\label{eq:Hamiltonian}
\end{equation}
where $t$ is the effective hopping between
the Ni and the Na atoms, $\varepsilon_{i}$
is the on-site energy, and 
the on-site Coulomb energies are $U_d$ and $U_s$ for the Ni (d)
and the Na (s) orbitals, respectively.

 The exchange interaction $J_{AF}$ between Ni atoms is obtained by calculating 
the ground state of the Hamiltonian Eq. \ref{eq:Hamiltonian} for
the ferromagnetic (FM) and antiferromagnetic (AF) configurations of
the Ni ($e_g^{(1)}$) spins and then taking the difference: 
$J_{AF}=E_{\uparrow\downarrow}-E_{\uparrow\uparrow}$.
For the FM case, we have two spin up electrons occupying the Na (s)
and
the Ni ($e_g^{(1)}$) states leading
to a three-dimensional configuration space. Similarly, for the AF case, we have a $9\times
9$ Hamiltonian (one spin up and one spin down electrons distributed among three orbitals
of either spin, so that $^3C_1 \times {^3C_1}$ = 9).  
The ground-state energies may be obtained by diagonalization of the Hamiltonians or by
using the fourth order 
perturbation theory.
The latter yields
\begin{equation}
J_{AF}  =  -\frac{2t^{4}}{\Delta_{sd}^{2}}\left[
\frac{2}{U_{s}+2\Delta_{sd}} +  \frac{1}{U_{d}}   \right],
\label{eq:Jc}
\end{equation}
where $\Delta_{sd}$ is the charge transfer energy from Ni (d) to Na (s).
 The first term is due to the exchange via the intermediate state with the two electrons on
the Na atom, while the second is due to hopping between the two Ni orbitals via the Na
atom. The exchange coupling  is clearly antiferromagnetic. Taking
typical values:
$t\approx 0.3$ eV, $\Delta_{sd}\approx5$ eV, $U_{d}\approx 5$ eV, and  
$U_{s}\approx2$ eV, the magnitude of the exchange is $J_{AF}\approx -0.1$ meV, which is the
right 
order of magnitude as compared to the experimental value of $-1$ K.

We thank Z.S. Popovic for valuable discussions and acknowledge financial support 
by the U.S. Department of Energy under Contact
No. DE-FG02-00ER45818.

%%%%%%%%%%%%% REFERENCES %%%%%%%%%%%%%%%%

\end{document}